\documentclass[twocolumn,aps,floatfix]{revtex4}
\usepackage{graphicx,epsfig,subfig,dcolumn,bm,mathrsfs,amsmath}
\usepackage{color}
\newcommand{\ud}{\mathrm{d}}

\begin{document}

\title{Application of Self-Consistent Field Theory to Self-Assembled Bilayer Membranes}

\author{Pingwen Zhang}
\email{pzhang@pku.edu.cn}
\affiliation{
 LMAM, CAPT and School of Mathematical Sciences,
 Peking University, Beijing 100871, P.R.~China}

\author{An-Chang Shi}
\email{shi@mcmaster.ca}
\affiliation{
 Department of Physics \& Astronomy, McMaster
 University, Hamilton, Ontario Canada L8S 4M1}


\begin{abstract}
Bilayer membranes self-assembled from amphiphilic molecules such as lipids, surfactants and block copolymers are ubiquitous in biological and physiochemical systems. The shape and structure of bilayer membranes depend crucially on their mechanical properties such surface tension, bending moduli and line tension. Understanding how the molecular property of the amphiphiles determine the structure and mechanics of the self-assembled bilayers requires a molecularly detailed theoretical framework. The self-consistent field theory provides such a theoretical framework, which is capable of accurately predicting mechanical parameters of self-assembled bilayer membranes. In this mini review we summarize the formulation of the self-consistent field theory, as exemplified by a model system composed of flexible amphiphilic chains dissolved in hydrophilic polymeric solvents, and its application to the study of self-assembled bilayer membranes. \\

\end{abstract}


\maketitle

\section{Introduction}
\label{sec:introduction}

Bilayer membranes self-assembled from amphiphilic molecules are important components of biological and physiochemical systems. In particular, bilayers self-assembled from lipids are ubiquitous in the cell, separating the cell from the exterior environment as well as encasing its internal organelle structures. Numerous biological functions are associated with the deformation, fusion and fission of lipid membranes \cite{BAlberts2014}. Amphiphilic molecules, or apmphiphiles, are molecules composed of hydrophilic and hydrophobic components. Examples of amphiphiles include lipids, surfactants and block copolymers. Due to the competing hydrophilic and hydrophobic interactions, amphiphilic molecules in aqueous solution spontaneously self-assemble into a variety of structures including spherical micelles, cylindrical micelles, and bilayers \cite{SASafran1994}. One extremely interesting and important self-assembled structure is the bilayer, in which the hydrophilic parts are located on the outer surface of the bilayer whereas the hydrophobic blocks are hidden in the interior of the membrane. At the molecular scale, the lateral arrangement of the molecules in a bilayer membrane is disordered, thus the membrane could be regarded as a two-dimensional fluid. At the mesoscopic scale, bilayer membranes could be regarded as two-dimensional surfaces embedded in the three-dimensional space. The unique structure of self-assembled bilayers results in several intriguing properties \cite{RLipowsky1998}: the bilayers can form closed membranes without edges, such as vesicles and cell walls; they are extremely flexible and highly deformable; and despite their flexibility, they keep their structural integrity even under strong deformations. The morphology formed by the self-assembled bilayer membranes is strongly affected by the property of the amphiphilic molecules. 

Phenomenologically, a bilayer membrane could be modelled as a two-dimensional surface. The formation and stability of membrane morphologies could then be understood by considering the mechanical property of this surface. Assuming the deformation from equilibrium state is small, the energy of a deformed bilayer membrane can be described by a linear elasticity theory, resulting in the celebrated Helfrich model of membranes \cite{WHelfrich1973,ZCOuYang1999}. Specifically the elastic energy of the Helfrich model of an open membrane is given by,
\begin{equation}
  \label{eq:Helfrich}
  F = \int \left[ \gamma+2\kappa_M (M-c_0)^2 + \kappa_G G \right] \ud A + \int \sigma \ud L,
\end{equation}
where $M=(c_1+c_2)/2$ and $G=c_1 c_2$ are the local mean and Gaussian curvatures of the deformed bilayer ($c_{1,2}$ are the two principal curvatures). The last term in Eq.\ (\ref{eq:Helfrich}) represents the edge energy of an open membrane, and it vanishes for closed membranes. $\gamma$ is the surface tension (surface energy per unit area) of the membrane. For a tensionless membrane $\gamma=0$. The energetics of a bilayer is specified by the elastic constants, $c_0$, $\kappa_M$, $\kappa_G$ and $\sigma$, corresponding to the spontaneous curvature, the bending modulus, the Gaussian modulus and the edge line tension, respectively. It should be noticed that the Helfrich free energy is a linear elastic model, which includes only the lowest-order contributions to the free energy from the curvatures. Higher-order contributions should be included if the curvature of the membrane is high. The mechanical property of the membrane within this surface model are characterized by a set of mechanical parameters, {\it i.e.}, the surface tension $\gamma$, spontaneous curvature $c_0$, bending modulus $\kappa_M$, and Gaussian modulus $\kappa_G$. For an open membrane, its energetic depends further on the line tension $\sigma$, which is the edge energy per unit length of the membrane. The elastic properties can be used to analyze and explain numerous phenomena associated with vesicle shape and deformation, membrane fusion, and other relevant membrane activities \cite{ZCTu2008,ZCTu2013,ZCTu2014}. The line tension of an open edge is a key parameter for understanding the processes of disc-to-vesicle transformation \cite{JFLi2013a}, vesicle-pore formation \cite{CLTing2011} and membrane fusion \cite{MMuller2002,KKatsov2004}.

In the study of bilayer membranes, one of the main objectives for experimentalists and theorists is to determine these elastic parameters and to understand their physical origins. In particular, a great number of theoretical and experimental studies have focused on the relationship between the molecular properties, such as composition, geometrical shape and interactions between lipid species, of the amphiphilic molecules and the  mechanical properties of the self-assembled membranes \cite{JFNagle2013}. Experimentally a  number of techniques have been developed for the measurement of bending moduli of lipid membranes \cite{JFNagle2013,RDimova2014,JFNagle2015}. A large body of knowledge about the elastic properties of lipid membranes has been obtained from these experiments, although uncertainties about these properties remain \cite{JFNagle2015}. Besides experimental measurements, simulation methods have been developed to obtain the elastic constants of bilayer membranes \cite{AJSodt2013,ZALevine2014}. Because free energy is not available directly from most of the simulation techniques, studying the elastic constants of membranes in simulations largely depends on either monitoring the thermal fluctuations of a membrane, or by integrating the local contributions from the pressure tensors. Furthermore, measuring the Gaussian modulus presents a challenge to simulations since the change of the Gaussian curvature contribution to the free energy of close membranes only occurs when there are topological changes of the system. 

An alternative strategy to obtain the elastic constants of membranes is to compare the free energy of membranes with different shapes of definite curvatures. By studying bilayers in different geometries ({\it e.g.} planes, cylinders, spheres), it is possible to obtain several elastic constants simultaneously. This approach requires an accurate computation of the free energy of the curved membranes, which can be implemented using different coarse-grained model of the amphiphilic molecules. Among the different theoretical frameworks developed for amphiphilic molecules, the self-consistent field theory (SCFT) provides a versatile platform for the study of self-assembled bilayer membranes. The self-consistent field theory is originally developed for the study of inhomogeneous polymeric systems \cite{ACShi2004,GHFredrickson2006}. It has been shown by numerous researchers in the past decades that the SCFT provides a flexible and accurate framework for the study of self-assembled structures included the ordered phases of multiblock copolymers \cite{ZGuo2008,WXu2013} and complex micelles \cite{JZhou2011,JZhou2014}. It has also been demonstrated that SCFT provide a rich free energy landscape for the study of transition pathways between different morphologies \cite{XCheng2010}. Furthermore, numerical calculations using the self-consistent field theory have been carried out to study the elastic properties of bilayers, using diblock copolymer/homopolymer blends as a coarse-grained model to mimic the amphiphile/solvent system \cite{MMuller2006,JFLi2013b,ADehghan2015}. In these theoretical studies, the focus has been on relating the elastic properties to the molecular parameters of the amphiphilic molecules.

In this paper, we review recent application of self-consistent field theory to the study of the mechanical properties of self-assembled bilayer membranes. The self-consistent field theory starts from a coarse-grained model of the amphiphilic molecules. The detailed formulation of the self-consistent field theory depends on the specificity of the molecular models. In this review we focus on a coarse-grained model  in which the molecule species are modelled as flexible polymers \cite{MMuller2006}. The advantage of this formulation of the self-consistent field theory is that the single-molecular partition function can be computed by solving a modified diffusion equation \cite{ACShi2004,GHFredrickson2006}. It should be noticed that the amphiphilic molecules can also be modelled using lattice models \cite{HPera2014} or rotational isomeric state models \cite{MJUline2012}. In the SCFT formulation, the bilayer membrane is described by a microscopic model of flexible amphiphilic chains (AB diblock copolymers) dissolved in hydrophilic solvent molecules (A homopolymers). Free energy of self-assembled bilayer membranes with specific shapes is obtained by solving the corresponding SCFT equations. The mechanical parameters of the membranes can be extracted by fitting the SCFT free energies of membranes with different geometries to an appropriate energy expression for the continuum elastic model \cite{JFLi2013b,ADehghan2015}.

The remainder of this review is organized as follows: Section~\ref{sec:model} describes the coarse-grained model of bilayer membranes, the self-consistent field theory framework and geometrical constraints used in the study. A brief review of recent results of the application of self-consistent field theory to elastic properties of self-assembled bilayer membranes is presented in section~\ref{sec:results}. Possible future application of the self-consistent field theory to the study of mechanics and morphology of self-assembled bilayer membranes is outline in section~\ref{sec:summary}.

\section{Coarse-Grained Model and SCFT Framework}
\label{sec:model}

In this section, we first described a coarse-grained model of self-assembling amphiphile/solvent system, in which the molecular species are modelled by flexible amphiphilic polymer chains, or AB diblock copolymers, dissolved in hydrophilic solvent molecules modelled as A homopolymers \cite{MMuller2006}. We then give a brief review of the self-consistent field theory formulated in grand canonical ensemble \cite{JFLi2013b,ADehghan2015}, which is used to compute the free energy of a self-assembled bilayer membrane. Finally we describe the implementation of SCFT in different geometries, which is used to extract the elastic constants.

\subsection{Model of Amphiphilic Molecules}
\label{sec:CGModel}

There are many types of amphiphilic molecules ranging from small molecular surfactants to macromolecular analogues such as diblock copolymers. Despite the variation of the molecular details, the self-assembly behaviour of the amphiphilic molecules are determined mainly by a few common features of the molecules, {\it i.e.}  the relative strength of the hydrophilic-hydrophobic interactions and the relative size of the two components. It has been demonstrated that coarse-grained model of amphiphilic molecules could provide useful information about the thermodynamic behaviour of the self-assembled bilayer membranes \cite{MMuller2006}. In this review we will focus on one particularly useful coarse-grained model in which the molecular species are modelled as flexible polymers \cite{JFLi2013b,ADehghan2015}. Although many molecular details have been ignored, the main features of the molecules are kept in the coarse-grained model. The advantages of the polymeric model is that there have been tremendous development in theoretical study of polymers in the last decades. The availability of sophisticated and efficient theoretical methods makes it possible to investigate accurate free energy of self-assembled bilayer membranes with different shapes and sizes.  

Specifically, the molecular species of the system are modelled by using corresponding polymeric components. In what follows we will use a system composed of one type of amphiphlic molecules (lipids) and water as a model system to describe the theoretical framework. For this simple system, the two molecular species are represented by AB diblock (amphiphilic) copolymers and A-type (hydrophilic) homopolymers. The self-assembly of the system is examined by studying the thermodynamic behaviour of binary mixtures of AB diblock copolymers and A-homopolymers in a volume $V$. Extension of this simple model to include electrostatic interactions and more complex block copolymer architectures \cite{CLTing2011}, as well as to multicomponent bilayer membranes \cite{ADehghan2015}, is straightforward.  

In the simplest polymeric model of an amphiphilic molecule, the hydrophilic and hydrophobic part is represented by the A and B blocks, which are connected at their ends to form a single diblock copolymer chain. The molecular parameters characterizing an AB diblock copolymer chain are its total length, or the degree of polymerization, $N$ and the lengths of the hydrophilic (A) and hydrophobic (B) blocks, $N_A$ and $N_B$ ($N_A+N_B=N$). Equivalently the composition of the amphiphile can be described by the volume fraction of the hydrophilic blocks in the amphiphile, $f_A=N_A/N$. For simplicity, we assume that both the copolymers and the homopolymers have equal chain length characterized by a degree of polymerization $N$. Furthermore, the A/AB blend is  assumed to be incompressible, and both monomers (A and B) have the same monomer density $\rho_0$ (or the hardcore volume per monomer is $\rho_0^{-1}$) and Kuhn length $b$. The interaction between the hydrophilic and hydrophobic monomers is described by the Flory-Huggins parameter $\chi$.

\subsection{Self-Consistent Field Theory}
\label{sec:freeE}

For the specific coarse-grained model or the AB/A mixtures, the task is to, first of all, demonstrate that self-assembly does occur in the model system leading to the formation of bilayer membranes. Then the free energy of the self-assembled membranes should be obtained so that the mechanical constants of the membrane could be extracted from the theory. The polymeric self-consistent field theory (SCFT) provides an ideal framework for this task. The SCFT of polymers was originally developed to study the thermodynamic behaviour of inhomogeneous polymer phases. Due the efforts of a large number of researchers, efficient methods to solve the SCFT equations have been developed, enabling the study of complex ordered phases of block copolymers as well as micellar morphologies of block copolymer solutions. One advantage of SCFT is that the theory provides an accurate calculation of the free energy of the system.

Details of the general theory of SCFT are found in a number of review articles and monographs \cite{ACShi2004,GHFredrickson2006}. In what follows we give a brief summary of the SCFT for AB/B blends formulated in the grand-canonical ensemble \cite{JFLi2013b}. In this formulation the chemical potential of the homopolymers is used as a reference. The controlling parameter is the copolymer chemical potential $\mu_c$, or its activity $z_c=\exp(\mu_c)$. This system mimics a model system composed of one type of lipid (the AB diblock copolymers) in aqueous solutions. 

The starting point of the SCFT is the partition function of the system which can be transformed into a field theoretical description, in which the participation function is written as a functional integral with the integrand specified by a free energy functional. Within the mean-field approximation, the grand free energy of a binary mixture is given by \cite{ACShi2004,GHFredrickson2006},
\begin{eqnarray}
  \frac{N \mathscr{F}}{k_B T \rho_0} &=& \int
  \ud \mathbf{r} \Big[ \chi N \phi_A(\mathbf{r}) \phi_B(\mathbf{r})
  - \omega_A(\mathbf{r}) \phi_A(\mathbf{r}) \nonumber \\
  &-& \omega_B(\mathbf{r}) \phi_B(\mathbf{r}) - \xi(\mathbf{r})
  (1-\phi_A(\mathbf{r})-\phi_B(\mathbf{r})) \nonumber \\
  &-& \psi \delta(\mathbf{r}-\mathbf{r}_1) (\phi_A(\mathbf{r})
  - \phi_B(\mathbf{r})) \Big] \nonumber \\
  \label{eq:freeE}
  &-& z_c Q_{c}-Q_{h} ,
\end{eqnarray}
where $\phi_{\alpha}(\mathbf{r})$ and $\omega_{\alpha}(\mathbf{r})$ are the spatially varying concentration and the meanfield of the $\alpha$-type monomers ($\alpha=A,B$), $\xi(\mathbf{r})$ is a Lagrange multiplier used to enforce the incompressibility condition. A second Lagrange multiplier, $\psi$, is introduced to ensure the formation of bilayers with different geometries. A delta function, $\delta(\mathbf{r}-\mathbf{r}_1)$, is used to ensure that the $\psi$ field only operates on the interface at a prescribed position $\mathbf{r}_1$. The last two terms in Eq.~(\ref{eq:freeE}) are the configurational entropies, which are specified by the single-chain partition functions for two types of polymers, $Q_{c}$ and $Q_{h}$.

For the copolymer, the partition function has the form $Q_{c} = \int \ud \mathbf{r} q_{c}(\mathbf{r},1)$, where $q_{c}(\mathbf{r},s)$ is an end-integrated propagator, and $s$ is a parameter that changes from 0 to 1 along the length of the polymer. The propagator is solution of the modified diffusion equation \cite{ACShi2004,GHFredrickson2006},
\begin{equation}
  \label{eq:mde}
  \frac{\partial}{\partial s} q_{c}(\mathbf{r},s) = R_g^2 \nabla^2 q_{c}(\mathbf{r},s) - \omega_{\alpha}(\mathbf{r}) q_{c}(\mathbf{r},s),
\end{equation}
where $R_g=b\sqrt{N/6}$ is the radius of gyration of the copolymer chains.
The mean field $\omega_{\alpha}(\mathbf{r})$ is a piece-wise function where $\alpha=A$ if $0<s<f_A$ and $\alpha=B$ if $f_A<s<1$.
The initial condition is $q_c(\mathbf{r},0)=1$. Since the copolymer has two distinct ends, a complementary propagator $q_c^{\dagger}(\mathbf{r},s)$ is introduced.
It satisfies Eq.~(\ref{eq:mde}) with the right-hand side multiplied by $-1$, and the initial condition $q_c^{\dagger}(\mathbf{r},1)=1$.
For the homopolymer, one propagator $q_h(\mathbf{r},s)$ is sufficient, and the single-chain partition function has the form $Q_h = \int \ud \mathbf{r} q_h(\mathbf{r},1)$.
The modified diffusion equations can be solved in real space or reciprocal space using a number of numerical methods developed previously \cite{GHFredrickson2006}.

\begin{figure}[htp]
  \includegraphics[width=1.0\columnwidth]{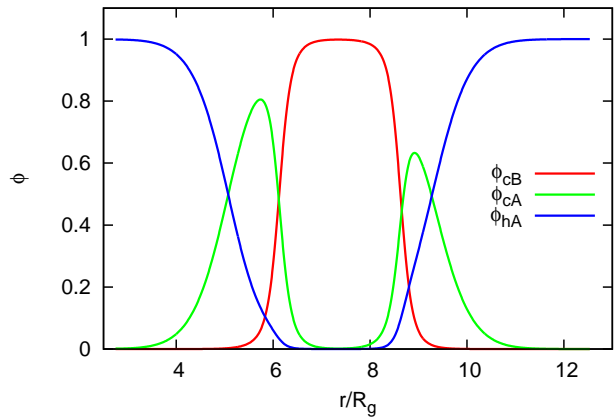}
  \caption{Density profile for a tensionless spherical bilayer with a radius $R=7.7R_g$, $f_A=0.40$ and $\chi N=30$. (Courtesy of J. Zhou)}
  \label{fig:profile}
\end{figure}

The SCFT method employs a mean-field approximation to evaluate the free energy using a saddle-point technique. The mean-field equations, or the SCFT equations, are obtained by requiring that the functional derivatives of the free energy (\ref{eq:freeE}) vanish,
\begin{equation*}
  \frac{\delta \mathscr{F}}{\delta\phi_{\alpha}} = \frac{\delta \mathscr{F}}{\delta\omega_{\alpha}} = \frac{\delta \mathscr{F}}{\delta\xi} = \frac{\delta \mathscr{F}}{\delta\psi} =0.
\end{equation*}
Carrying out these functional derivatives results in the following SCFT equations:
\begin{eqnarray*}
  \phi_A(\mathbf{r}) &=& \int_0^1 \ud s \, q_h(\mathbf{r},s) q_h(\mathbf{r},1-s) \nonumber \\
  && + z_c \int_0^{f_A} \ud s \, q_c(\mathbf{r},s) q_c^{\dagger}(\mathbf{r},s), \\
  \phi_B(\mathbf{r}) &=&  z_c \int_{f_A}^1 \ud s \, q_c(\mathbf{r},s) q_c^{\dagger}(\mathbf{r},s), \\
  \omega_A(\mathbf{r}) &=& \chi N \phi_B(\mathbf{r}) + \xi(\mathbf{r}) - \psi \delta(\mathbf{r}-\mathbf{r}_1),\\
  \omega_B(\mathbf{r}) &=& \chi N \phi_A(\mathbf{r}) + \xi(\mathbf{r}) + \psi \delta(\mathbf{r}-\mathbf{r}_1),\\
  1 &=& \phi_A(\mathbf{r})+\phi_B(\mathbf{r}),\\
  \phi_A(\mathbf{r}_1) &=& \phi_B(\mathbf{r}_1).
\end{eqnarray*}
The SCFT equations are a set of nonlinear and nonlocal equations. Solving the SCFT equations usually requires numerical methods. In the past years a number of efficient and stable numerical techniques have been developed for the solution of the SCFT equations \cite{ACShi2004,GHFredrickson2006}. Many of these numerical methods could be used in finding solutions of the SCFT equations corresponding to bilayer membranes.  A typical solution for a self-assembled bilayer membrane in the spherical geometry, corresponding to a vesicle, is shown in Figure~\ref{fig:profile}.

\begin{figure}[htp]
   \centering
   \includegraphics[width=1.0\columnwidth]{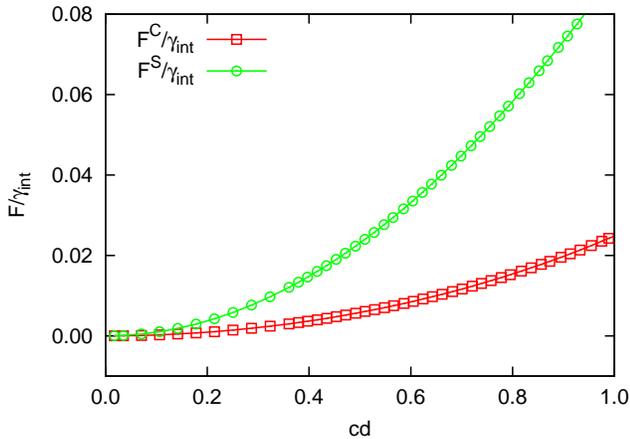}
   \caption{Excess free energy of a tensionless bilayer in the cylindrical (squares) and spherical (circles) geometries. $c=1/r$ and $d$ is the thickness of the flat bilayer. The parameters used in the calculations are $f_A=0.40$ and $\chi N=30$.
    (Courtesy of J. Zhou)}
   \label{fig:freeE}
 \end{figure}

For the study of elastic properties of bilayer membranes, the quantity of interest is the free energy of a system containing a bilayer membrane. The reference state of the system is an homogeneous mixture of the the copolymers and homopolymers. The SCFT equations for a homogeneous phase can be solved analytically. In particular, the bulk copolymer concentration $\phi_{\rm bulk}$ can be obtained as a function of the copolymer chemical potential $\mu_c$,
\begin{eqnarray*}
  \mu_c &=& \ln \phi_{\rm bulk} - \ln ( 1- \phi_{\rm bulk}) \nonumber \\
  &+& \chi N (1-f_A) \big[ 1-2(1-f_A)\phi_{\rm bulk} \big].
\end{eqnarray*}
The free energy of the homogeneous phase, $\mathscr{F}_{\rm bulk}$, can be obtained by inserting the homogeneous solution to the SCFT free energy, Eq.~\ref{eq:freeE}. For a bilayer membrane, its excess free energy ($\mathscr{F}-\mathscr{F}_{\rm bulk}$) is proportional to the area of the membrane. It is useful to define an excess free energy density, $F$, as the free energy difference between the systems with and without the bilayer membrane, divided by the area, $A$, of the membrane,
\begin{equation}
  \label{eq:F_excess}
  \Delta F = \frac{N(\mathscr{F}-\mathscr{F}_{\rm bulk})}{k_B T \rho_0 A}.
\end{equation}
An example of the excess free energy (in unit of $\gamma_{int}=\sqrt{\chi N/6}k_{B}Tb/ \sqrt{N}$) of cylindrical and spherical membranes as a function of the dimensionless curvature $cd$ ($d$ is the thickness of the bilayer) is shown in Figure~\ref{fig:freeE}. 
A comparison of this SCFT free energy with the Helfrich elastic energy (Eq.~\ref{eq:Helfrich}) can then be used to determine the various elastic constants.

\subsection{Geometrical Constraints}
\label{sec:geometry}

The central task of studying the mechanical property of bilayers is to obtain solutions of the SCFT equations corresponding to self-assembled membranes with given shape and size. The elastic constants can then be extracted by comparing the free energy with the Helfrich model (Eq.~\ref{eq:Helfrich}). In order to extract information on the various elastic properties, the excess free energy of a bilayer with different geometries need to be computed. In most of the studies, solutions of the SCFT equations corresponding to the five geometries have been obtained. These five geometers are (i) an infinite planar bilayer membrane, (ii) a cylindrical bilayer membrane with a radius $r$, which is extended to infinity in the axial direction, (iii) a spherical bilayer with a radius $r$, (iv) an axially symmetric disk-like membrane patch with a radius $R$, and (v) a planar membrane with a circular pore of radius $R$. The first three geometries, which can be reduced to a one-dimensional problem by using an appropriate coordinate system, are employed to extract the bending modulus and the Gaussian modulus as well as higher-order curvature moduli. Whereas the last two geometries, which are two-dimensional systems due to their axial symmetry, are used to extract the line tension of the membrane edge.

Bilayers with a given geometry can be stabilized by applying appropriate constraints to the SCFT equations. With these geometric constraints, the excess free energies for the five geometries, which are denoted $F^{X}$, where $X=0,C,S,D,P$ for the planar, the cylindrical, the spherical, the disk and the pore geometry, respectively, can be calculated. In general, the surface tension of the membrane depends on the chemical potential of the copolymers. A tensionless membrane is obtained when the chemical potential of the amphiphilic diblock copolymers, $\mu_c$, is adjusted carefully such that the grand free energy of the bulk system and of a planar bilayer are identical, \emph{i.e.}, $F^0 = 0$.

For a bilayer of finite thickness, the definition of the interface position or the membrane surface involves a certain degree of arbitrariness. To be consistent with our constraint method, the bilayer interface is defined by the mid point of the two positions where the A- and B-segment concentrations are equal. For a cylindrical membrane with radius $r$, the mean curvature is $M=1/(2r)$ and the Gaussian curvature vanishes. For a spherical membrane with radius $r$, the mean curvature and the Gaussian curvature are $M=1/r$ and $G=1/r^2$, respectively. With these curvatures, the Helfrich free energy (Eq.~\ref{eq:Helfrich}) is written in terms of the curvature $c=1/r$ in the cylindrical and spherical geometries,
\begin{eqnarray}
  \label{eq:fC_fit}
  F^C &=& -2 \kappa_M c_0 c + \frac{\kappa_M}{2} c^2 , \\
  \label{eq:fS_fit}
  F^S &=& -4 \kappa_M c_0 c + (2\kappa_M + \kappa_G) c^2,
\end{eqnarray}
These free energy expressions are to be compared with the SCFT free energy obtained by solving the SCFT equations for cylindrical and spherical bilayer membranes. In particularly the elastic constants $c_{0}$, $\kappa_{M}$ and $\kappa_{G}$ are obtained by fitting the SCFT free energy to these expressions. The SCFT free energy of the different bilayer membranes is obtained by inserting the SCFT solutions to the expression of the free energy. An example of the excess free energy is shown in Figure~\ref{fig:freeE}. These free energy curves are well described by the Helfrich expression (Eq.~\ref{eq:fC_fit}) for membranes with small curvature. Due to the symmetry of the bilayers the spontaneous curvature of the membranes is zero in this case. On the other hand the bending and Gaussian modulus, $\kappa_{M}$ and $\kappa_{G}$, are obtained by fitting the SCFT excess free energy to the Helfrich model.

\begin{figure}[htp]
  \includegraphics[width=1.0\columnwidth]{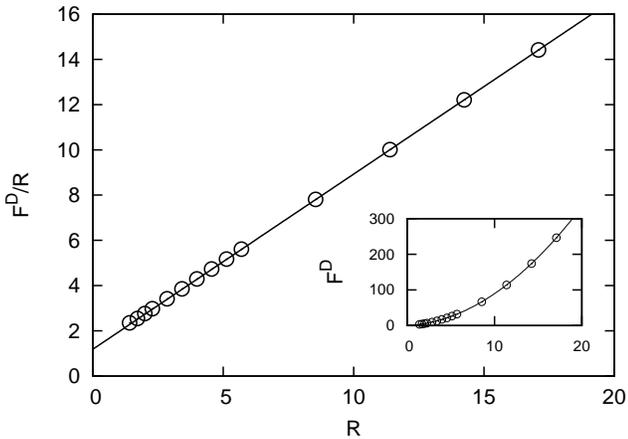}
  \caption{Plot of $\Delta F/R$ as a function of $R$ for a circular bilayer disk. The parameters used in the plot are $f_A=0.40$, $\chi N=30$ and $\mu = 4.4$. Inset: Plot of $\Delta F$ as a function of $R$. (Courtesy of Dr. J. Zhou)}
  \label{fig:line}
\end{figure}

The free energy of an open membrane contains contributions from the edge of the bilayers in the form of edge energy or line tension. For a flat circular disk with a large radius the excess free energy can be written as contributions from the surface tension and the edge energy, 
\begin{equation}
  \label{eq:fd}
  \Delta F = \pi\gamma R^2 + 2\pi \sigma R,
  \end{equation}
where $R$ is the radius of the disk. A typical plot of the excess free energy as a function of the disk radius is shown in the insect Fig.~\ref{fig:line}, demonstrating the parabolic feature of the function. For the purpose of obtaining the surface tensions $\gamma$ and line tension $\sigma$, we can plot the the quantity $\Delta F/R$ as a function of $R$ (Fig.~\ref{fig:line}), which is a linear function of $R$. The slope and interception of this linear curve can then be used to determine the surface tension ($\gamma$) and line tension ($\sigma$).

\section{Recent Applications: Elastic Constants of Self-Assembled Bilayers}
\label{sec:results}

In this section we give a brief review on recent applications of SCFT to the study of elastic properties of self-assembled membranes. The application of SCFT to the study of self-assembled bilayer membranes has been made by a number of researchers \cite{MMuller2006,JFLi2013b,ADehghan2015,CLTing2011}. While there have been significant contributions to this topic by a number of research groups, we will restrict the scope of the applications to our own work \cite{JFLi2013b,ADehghan2015}. The basic idea here is not to give an exhaustive review of the topic, rather we would like to use a few examples to illustrate the usefulness of the SCFT framework.


\subsection{Elasticity of Membranes: Bending and Gaussian moduli}

The linear elasticity theory of membranes (Eq.~\ref{eq:Helfrich}) has been used as a foundation for the discussion of many membrane properties including the shape and deformation of vesicles. In the original formation of the the elasticity theory, the elastic constants (spontaneous curvature and bending and Gaussian moduli) are phenomenological parameters of the model. These elastic constants are to be determined from calculations based on microscopic theories or to be measured from experiments. The self-consistent field theory provides an ideal theoretical framework for the determination of these elastic constants from a molecular model. In the past decades, a number of SCFT studies have devoted to the study of membrane elasticity. In a recent publication Li {\em et al.} \cite{JFLi2013b} reported a detailed SCFT study of the the elastic properties of a bilayer, including the bending modulus and the Gaussian modulus, by applying the self-consistent field theory to a coarse-grained model composed of amphiphilic chains dissolved in hydrophilic solvent molecules. Furthermore they have examined the region of validity of the the linear elasticity theory for the given coarse-grained models by computing the higher-order contributions to the free energy. 

Specifically the coarse-grained model consists of amphiphile/solvent mixtures modelled as blends of AB diblock copolymers and B homopolymers. Solutions of the SCFT equations corresponding to bilayer membranes in five different geometries are obtained: the planar, the cylindrical, the spherical, the disk, and the pore geometry. The corresponding free energy of membranes is used to extract the elastic constants of the membrane. The effects of molecular properties, such as the hydrophilic volume fraction $f_A$ and the interaction parameter $\chi$, on the elastic constants are investigated. The general trend from study is that: (i) the bending modulus $\kappa_{M}$ is a concave function of the hydrophilic volume fraction $f_{A}$ and it is with a maximum around $f_A=0.50$, however the $f_{A}$-dependence is weak; (ii) the Gaussian modulus $\kappa_{G}$ is a monotonically decreasing function of $f_{A}$, the ratio of the Gaussian modulus and the bending modulus changes from 1 to -2 as $f_A$ is increased from 0.30 to 0.70. ; (iii) the value of $\kappa_{M}$ increases as $\chi N$ is increased while the value of $\kappa_{G}$ depends weakly on $\chi N$. 

Furthermore Li {\it et al.} \cite{JFLi2013b} also investigated the region of validity of the linear elasticity theory and concluded that the linear elasticity model is accurate at small membrane curvatures, but it becomes inaccurate when the radius of curvature is comparable to the membrane thickness. Higher-order curvature moduli have been introduced and calculated in their study.

\subsection{Line Tension of Open Membranes}

Line tension or the edge energy of open membranes is an important parameter of the membranes. In particular, the competition between the bending energy and edge energy determines whether a finite bilayer membrane would form a flat disk or an enclosed vesicle. The line tension also plays an important role in the formation of pores in membranes. Therefore accurate determination of line tension of membranes is an important application of the self-consistent field theory. 

In the recent study of Li {\it et al.} \cite{JFLi2013b}, the line tension of self-assembled membranes has been studied by applying the SCFT to the coarse-grained model. Specifically SCFT solutions corresponding to an open disk or pore have been obtained. The free energy of the disk or pore as a function of their size $R$ is then used to extract the line tension of the membranes. For tensionless membranes, these authors found that as the hydrophilic volume fraction ($f_{A}$) increases, the line tension $\sigma$ decreases linearly from a positive value to zero and even to a negative value, indicating that lipids with large head groups can be used as edge stabilizers.  

Very recently, Dehghan {\it et al.} \cite{ADehghan2015} have investigated the line tension or edge energy of self-assembled multicomponent bilayer membranes using the self-consistent field theory. The coarse-grained model consists of lipid species modelled as amphiphilic AB/ED diblock copolymers in a solvent, modelled as C-homopolymers. Solutions of the SCFT equations are obtained for a bilayer membrane with a pore of fixed size. By fitting the SCFT free energy of the pore to the elasticity model (Eq.~\ref{eq:fd}), the line tension of the membrane can be obtained. The effects of the composition, geometrical shape and interactions of lipid species on the line tension were examined. Their results can be summarized as follows. For a bilayer system composed of cylindrical, cone- and inverse cone-shaped lipid species, an increase in the concentration of the cone-shaped lipids results in a decrease in the line tension. In contrast to the behaviour of cone-shaped lipids, an increase in the concentration of the inverse cone-shaped molecules results in an increase in the pore line tension. The mechanism underlying the decrease or increase in the line tension could be attributed to the segregation of the molecules within the membrane. Furthermore, an increase in the repulsive interaction between the head groups results in a decrease in the pore line tension.

\section{Conclusions and Outlook}
\label{sec:summary}

In this paper we provide a brief review of the self-consistent field theory (SCFT) and its application to the study of self-assembled bilayer membranes. The SCFT is a flexible and accurate theoretical framework for the study of inhomogeneous morphologies formed from complex molecular systems. It has been well documented that the application of SCFT to polymeric systems containing block copolymers has resulted in a good understanding of the principles of nanostructure formation in that system \cite{ZGuo2008,WXu2013}. It has been demonstrated that the application of SCFT to coarse-grained model of amphiphilic molecules could lead to insights about the relationship between molecular properties of the amphiphiles and the mechanical properties of the self-assembled bilayer membranes. Although the coarse-grained model ignores many molecular details of lipids, the results and conclusion from the SCFT study provides a qualitative understanding of the interplay between the lipid properties and mechanics of membranes. 

The starting point of SCFT framework is the construction of coarse-grained models of the amphiphilic molecules. The detailed formulation of the self-consistent field theory depends on the specificity of the coarse-grained models. In order to take advantage of the sophisticated techniques developed for polymeric SCFT, it is useful to model the molecule species as flexible polymers. Specifically, the simplest coarse-grained model of amphiphilic molecular systems consists of flexible amphiphilic chains (AB diblock copolymers) dissolved in hydrophilic solvent molecules (A homopolymers). Accurate free energy of self-assembled bilayer membranes with specific shapes can be obtained by solving the corresponding SCFT equations. The availability of the free energy allows the computation of the mechanical parameters of the membranes by fitting the SCFT free energies of membranes with different geometries to an appropriate energy expression for the continuum elastic model. Recent studies of the polymeric coarse-grained models have revealed interesting relationship between the mechanical properties of the bilayer membranes and the molecular properties. In particular, SCFT results have revealed how the segregation of amphiphilic molecules could be used to regulate the line tension of open membranes \cite{ADehghan2015}. Furthermore, the SCFT could also be used to establish the region of validity of the linear elasticity theory \cite{JFLi2013b}.

In this review we focused on the study of mechanical properties of bilayer membranes using SCFT applied to five bilayer geometries. It should be emphasized that the application of SCFT to bilayer membranes is not restricted to the study of simple geometries as reviewed here. For example, SCFT has been used to examine morphological changes of membranes such as membrane fusion \cite{MMuller2002} and pore formation \cite{CLTing2011}, revealing mechanisms underlying pore formation of membranes under tension and fusion of two apposed membranes. Furthermore, extension of the theory to include more complex molecular architectures and electrostatic interactions is straightforward \cite{CLTing2011,ACShi1999}. 

Despite the success of applying SCFT to the study of self-assembled bilayer membranes, a number of theoretical challenges remain. First of all, the coarse-grained model of flexible polymers could be improved to include more molecular details of the amphiphilic molecules. For example, proper description of the orientational order of the lipids tails would require extending the flexible polymeric model to semiflexible chain model. When a non-Gaussian polymer model is used, the SCFT field theory should be extended to properly account the conformational statistics and the SCFT equations should be modified accordingly. Secondly many membrane configurations may be metastable or even unstable, thus requiring an extension of the theory to non-equilibrium structures. For example, the lipid composition of the inner and outer leaflets of many biological membranes are different. The asymmetric composition arises naturally in cell membranes when the lipids are generated and initially inserted into the inner leaflet, and the asymmetry often persists over long periods of time because the spontaneous flip-flop of lipids between two leaflets is extremely slow \cite{SSanyal2009}. The current SCFT model in general deals with thermodynamical equilibrium thus it is challenging to treat dynamic processes. Techniques exploring the free energy landscape \cite{CLTing2011,XCheng2010} could provide useful information for the understanding of morphological transitions of membranes. Thirdly, many biological processes involve protein-membrane interactions \cite{BAlberts2014}. It is therefore desirable to develop coarse-grained models for proteins and membranes. Application of SCFT to coarse-grained model of proteins and lipids could lead to insights on how different proteins affect the curvature and structure of membranes. Finally, the computation of the elastic constants is done by fitting the excess free energy to the Helfich model. It would be useful to derive the Helrich model from the SCFT theory, so that these elastic constants could be calculated from the knowledge of the SCFT solutions of the bilayers. Such a theoretical framework would also provide useful insights into the factors affecting the elasticity of the bilayers.

 \begin{acknowledgments}
We are grateful to Jiajia Zhou, Ashkan Dehghan and Jianfeng Li for many valuable discussions. 
We acknowledge support from the National Science Foundation of China (Grant No.\, 11421101, 21274005) and the Natural Sciences and Engineering Research Council (NSERC) of Canada.
 \end{acknowledgments}




\end{document}